\documentclass[12pt]{article}

\usepackage{graphicx}
\usepackage{amssymb}
\usepackage[applemac]{inputenc}
\title{Essential spectrum in vibrations of thin shells in membrane approximation.  Propagation of singularities}
\author{Alain CAMPBELL \thanks{Groupe de Mécanique, Modélisation 
Mathématique et Numérique,
Laboratoire de Mathématiques Nicolas Oresme, UMR 6139.
BP 5186 Université de Caen, 14032 Caen Cedex. France}}
\begin{document}

\maketitle

\centerline{ABSTRACT}

The spectral problem of thin elastic shells in membrane approximation does not satisfy
the classical properties of compactness and so there exists an essential spectrum.
In the first part, we propose to determinate this spectrum and the weakness directions
in the shell. We particularly study the case of homogeneous and isotropic shells with 
some examples. In the second part, we consider an elementary model problem to study the
propagation of singularities and their reflections at the boundary of the domain. In the last, 
we study the problem of propagation for an isotropic cylindrical shell and we show that
the equation of propagation does not depend on the Poisson coefficient.\\
{\it keywords : shell, essential spectrum, propagation of singularities} 

\section{Introduction}

\subsection{Classical and non classical vibrating problems in shell theory}

We consider a thin shell with a middle surface $S$. This surface is described by the map,
\begin{equation}
y=(y^1,y^2)\in \Omega \rightarrow r(y^1,y^2)\in \mathbb{R}^3
\end{equation}
where $\Omega$ is a domain of the plane. Let $u(y)=(u_1(y),u_2(y),u_3(y))$ be the displacement vector of the surface and its covariant components.
We introduce the Hilbert space $H=(L(\Omega))^3$ and we denote by $(u,v)$ the scalar product. 
The displacement
$u$ belongs to the subset $V_1= H^1(\Omega) \times H^1(\Omega)\times H^2(\Omega)$ which can be modified
 to take boundary conditions in account. The
variational form of the problem of vibrations is (cf.[12, ch. VI]),

We search for $u \in V_1$ so that,
\begin{equation}
\forall v \in V_1,\ a_m(u,v) + \epsilon^2 a_f(u,v) = \lambda (u,v)
\end{equation}
The bilinear 
forms $a_m$ and $a_f$ correspond respectively to the membrane problem and the flexion problem. They
are continuous on $V_1$.
We denote by $\lambda$ the spectral parameter. 
This problem is classical with a selfadjoint operator and compact resolvent and so 
there exists a sequence of eigenvalues
 \begin{equation}
O<\lambda_0≤ \lambda_1≤...≤\lambda_k≤  \longrightarrow + \infty
\end{equation}
with orthogonal modes (cf.[12, ch.I]).

If the relative thickness of the shell, $\epsilon$, is very small, then the membrane approximation is an appropriate representation. 
The formulation of this problem is different. In this case, $u$ belongs to the space $V= H^1(\Omega) \times H^1(\Omega)\times L^2(\Omega)$.
The inclusion of $V$ in $H$ is dense  and continuous but is not compact. The problem is written as,

We search for $u \in V$ so that,
\begin{equation}
\forall v \in V,\ \ a_m(u,v)  = \lambda (u,v)
\end{equation}
This spectral problem is an elliptic system with mixed order. The classical properties of
compactness are not satisfied and the spectrum both contains a sequence of eigenvalues depending
on the domain, and an essential spectrum. 

\subsection{Essential spectrum}

Let $H$ be a Hilbert space and $A$ a selfadjoint operator. The resolvent set is defined by,
\begin{equation}
\rho(A) = \{ \ \zeta \ / \  (A- \zeta Id)^{-1} \in {\cal L}(H) \  \}
\end{equation}
Its complement, the spectrum $\Sigma(A)$, is constituted,

\parindent =30pt - of isolated eigenvalues of finite multiplicity; for these $\zeta$, 
$(A- \zeta Id)^{-1}$ does not exist.

- of other values for which $(A- \zeta Id)^{-1}$ exists but does not belong
to ${\cal L}(H)$. 
They are eigenvalues of infinite multiplicity, accumulation points of eigenvalues
and continuous spectrum.

\parindent=0pt
The set of these $\zeta$ which are not isolated eigenvalues of finite multiplicity is the essential spectrum $\Sigma_{ess}(A)$.

It can be characterized as the set of $\zeta$ for which there exists
a sequence $(u_k)$ called Weyl's sequence so that,
\begin{equation}
\begin{array}{rl}
\mid \mid u_k \mid\mid &= 1\\
u_k & \longrightarrow 0 \ \hbox{ in H weakly}\\
(A-\zeta Id)(u_k ) &\longrightarrow 0 \ \hbox{ in H strongly}
\end{array}\end{equation}
For very small data, we can obtain a large response and so these sequences can be physically interpreted 
as some kind of resonance. This local phenomena are quick oscillations in some 
directions which are called weakness directions.  We will see that the singularities will propagate 
along directions which are orthogonal to these weakness ones.

In the first part of this paper, we study the essential spectrum of a vibrating shell in membrane approximation. 
We show how it is possible to determine this set and the corresponding weakness directions and we give some examples.
In the second part, we introduce a non classical model problem and its essential spectrum and we investigate 
the propagation and reflection of singularities. Finally, in the last part, we study these problems for a vibrating cylindrical shell.

\section{The case of shells in membrane approximation}

Let $S$ be the middle surface of a thin shell, described by the map,
\begin{equation}
(y^1,y^2)\in \Omega \rightarrow r(y^1,y^2)\in \mathbb{R}^3
\end{equation}
We define the fundamental forms,
\begin{equation}
A(x^1,x^2) = a_{\alpha \beta}x^{\alpha}x^{\beta}
\end{equation}
and
\begin{equation}
B(x^1,x^2) = b_{\alpha \beta}x^{\alpha}x^{\beta}
\end{equation}
The equations of the vibrating shell in the membrane approximation give the following spectral problem,

\begin{equation}  \label {10}
\left\{ 
\begin{array}{cc}
-D_1 T^{11}-D_2 T^{21} &= \lambda u^1 \\
-D_1 T^{12}-D_2 T^{22} &= \lambda u^2 \\
-b_{\alpha \beta}T^{\alpha \beta} &=  \lambda u^3
\end{array} \right.
\end{equation}
where $u^1$, $u^2$ and $u^3$ are the contravariant components of the displacement 
and $T^{\alpha \beta}$, those of the stress tensor (cf.[11, ch. X], [1]).\
The covariant derivatives of tensor $T^{\alpha \beta}$  are given by,
\begin{equation}
D_{\alpha}T^{\alpha \beta}= {\partial T^{\alpha \beta} \over \partial y^{\alpha}}
+ \Gamma_{\alpha \delta}^{\alpha}T^{\delta \beta}
+ \Gamma_{\alpha \delta}^{\beta}T^{\alpha \delta }
\end{equation}
using the Christoffel symbols. \
We define the strains of the shell by coefficients $ \gamma_{\lambda \mu}(u)$,
\begin{equation} \gamma_{\lambda \mu}(u) = {1 \over 2}(D_{\lambda}u_{\mu}+D_{\mu}u_{\lambda})-b_{\lambda \mu}u_3
\end{equation}
where $D_{\lambda}$ denotes covariant derivative on $(S)$
and the behaviour of the shell is obtained with the elasticity coefficients $a^{\alpha \beta\lambda \mu}$ so that,
\begin{equation} \label {13}
T^{\alpha \beta}=a^{\alpha \beta\lambda \mu} \gamma_{\lambda \mu}(u)
\end{equation}
We will note ${\cal A} = [a^{\alpha \beta\lambda \mu}]$ the stiffness matrix. \\
By replacing the expressions (\ref{13}) of $T^{\alpha \beta}$ in (\ref{10}), we obtain an explicit  spectral problem on the displacement $u$. There appears derivatives of second order in $u_1$ and $u_2$ and of first 
order in $u_3$. The classical properties of compactness are not satisfied and there exists an essential spectrum.
We have weakness directions noted by $(\xi_1, \xi_2)$ and the orthogonal ones $(x_1,x_2)$
will be the directions of propagation of the singularities. The values of $\lambda$ and $(\xi_1, \xi_2)$
correspond to the non-ellipticity of the system in the Douglis and Nirenberg sense (cf.[11 ch. III] [5]). 
In this spectral problem, we have three equations corresponding to (\ref{10}). The equation $i$ is of order $(s_i+s_j)$
in the variable $u_j$ with $s_1=s_2=1$ and $s_3=0$.\\
To determine  $\lambda$ and $(\xi_1, \xi_2)$, we write that they 
are solutions of the equation we obtain by writing that the determinant of the principal symbol is zero:
\begin{equation}
det \pmatrix{a^{\alpha 1\beta 1}\xi_{\alpha}\xi_{\beta} &a^{\alpha 1\beta 2} \xi_{\alpha}\xi_{\beta} & 
 ia^{\alpha 1\zeta \eta}b_{\zeta \eta}\xi_{\alpha}   \cr
a^{\alpha 1\beta 2}\xi_{\alpha}\xi_{\beta} &a^{\alpha 2\beta 2} \xi_{\alpha}\xi_{\beta} & 
 ia^{\alpha 2\zeta \eta}b_{\zeta \eta}\xi_{\alpha}  \cr
ia^{\alpha 1\zeta \eta}b_{\zeta \eta}\xi_{\alpha}& ia^{\alpha 2\zeta \eta}b_{\zeta \eta}\xi_{\alpha} &
\lambda-a^{\alpha \beta \zeta \eta}b_{\alpha \beta}b_{\zeta \eta} \cr} =0 
\end{equation}
which we write in the condensated form, 
\begin{equation}
det \pmatrix{A_{11} & A_{12} & A_{13}   \cr
A_{12} & A_{22} & A_{23}  \cr
A_{13}& A_{23} &
\lambda-B_{33} \cr} =0 
\end{equation}
Whe then  have,
\begin{equation}
\lambda = B_{33} + {A_{13}^2A_{22}+A_{23}^2A_{11}-2A_{12}A_{23}A_{13} \over A_{11}A_{22}-A_{12}^2}
\end{equation}
The numerator $B_{33}(A_{11}A_{22}-A_{12}^2) +A_{13}^2A_{22}+A_{23}^2A_{11}-2A_{12}A_{23}A_{13}$  is polynomial of degree
4, homogeneous in $\xi_1, \xi_2$. By calculating every coefficient, we see that all of them are multiple of $det {\cal A}$ and finally
that it is exacty equal to $ det {\cal A} \ .\ [B(-\xi_2,\xi_1)]^2$.
It is easy to calculate the denominator,
\begin{equation}
A_{11}A_{22}-A_{12}^2=c_{22}\xi_1^4+c_{11}\xi_2^4+(c_{33}+2c_{12})\xi_1^2\xi_2^2-2c_{23}\xi_1^3\xi_2-2c_{13}\xi_1\xi_2^3
\end{equation}
where $c_{\alpha \beta}$ are the cofactors of matrix ${\cal A}$. \
We then obtain the relation between $\lambda$ and $(\xi_1, \xi_2)$. By replacing the components of a vector of the weakness direction 
by those of the direction of propagation $(x_1,x_2)$, we have,
\begin{equation}
(x_1,x_2)=(-\xi_2,\xi_1)
\end{equation}
and then,
\begin{equation}\label{20}
\lambda ={[B(x_1,x_2)]^2 \over 
s^{\alpha \beta \lambda \mu} x_{\alpha} x_{\beta}x_{\lambda} x_{\mu} }
\end{equation}
where $s^{\alpha \beta \lambda \mu}$ are the coefficients of ${\cal A}^{-1}$ (compliance coefficients).

Let us recall that the coefficients which appear in that last expression depend on $(y^1,y^2)$. If a point is given
on $S$, then the spectral parameter $\lambda$ belongs to a segment
\begin{equation}
\Sigma_{ess}=[\Lambda_1, \Lambda_2]
\end{equation}
and the whole essential spectrum is the set of all these segments when $(y^1,y^2)$ draw  $\Omega$.\\
By noting that $\lambda$ is a quotient of two quadratic forms in 
$X= (x_1^2,x_2^2,x_1x_2)$, 
\begin{equation}
\lambda ={ ^tXQ_1X \over   ^tXQ_2X}
\end{equation}
we can write that,
\begin{equation}
\Sigma_{ess}=[\Lambda_1, \Lambda_2] \subset [min\ (eigenvalue\ of\ Q_2^{-1}Q_1), max(eigenvalue\ of\ Q_2^{-1}Q_1)]
\end{equation}
and we obtain the classical inclusion (cf.[11 ch. XI]),
\begin{equation}
\Sigma_{ess} \subset [0, B_{33}]=[0, a^{\alpha \beta \zeta \eta}b_{\alpha \beta}b_{\zeta \eta}]
\end{equation}
We note that $0$ is reached in every hyperbolic point of the shell but not in elliptic ones as it is easily seen from (\ref {20}).\
Moreover, there are some cases for which $B_{33}$ is not reached. For instance, if we consider an isotropic cylindrical shell
($x^1= Ry^1; \ x^2 = Rcosy^2; \  x^3 = R siny^2$), then we have
\begin{equation}
\Sigma_{ess} = [0, {E \over R^2}]≠[0, B_{33}={E \over R^2 (1-\nu^2)}]
\end{equation}
Let us now consider the case of an isotropic shell. The elasticity coefficients are given by,
\begin{equation}
a^{\alpha \beta \lambda \mu}= {E \over 2(1+\nu)}\ \big(a^{\alpha  \lambda }a^{ \beta \mu}+a^{\alpha  \mu }a^{ \beta \lambda}+
{ 2\nu \over (1-\nu)}\ a^{\alpha \beta}a^{ \lambda \mu}\big
)\end{equation}
and so we have,
\begin{equation}
det {\cal A} = {E^3 \over 2(1-\nu^2)(1+\nu)}\ (a^{11}a^{22}-(a^{12})^2)
\end{equation}
and
\begin{equation} 
s^{\alpha \beta \lambda \mu} x_{\alpha} x_{\beta}x_{\lambda} x_{\mu}  =
{1 \over E}\ (a_{11}x_1^2 + a_{22}x_2^2+2a_{12}x_1x_2)^2
\end{equation}
where $[a_{\alpha \beta}] = [a^{\alpha \beta}]^{-1}$. Finally, we obtain the following 
outstanding form of (\ref {20}),
\begin{equation}
\lambda = E   \Big[\ {B(x_1,x_2) \over 
A(x_1,x_2) }\  \Big]^2
\end{equation}
where $A$ et $B$ are the two fundamental forms of the surface $S$. 
It appears that the essential spectrum depends only on
the geometry and the Young modulus but is independent of the Poisson
coefficient.
A geometrical interpretation of the quotient of the fundamental forms
${B(x_1,x_2) \over 
A(x_1,x_2) } $ is the normal curvature $k_x$ of the surface in direction $(x_1,x_2)$.
The essential spectrum is then exactly the segment,  
\begin{equation} 
\Sigma_{ess} = [\ E.Infk_x^2, \ E.Supk_x^2 \ ]
\end{equation}
Conversely, if  $\lambda \in \Sigma_{ess}$ is given, then we can find the couples $(x_1,x_2)$
which define the directions of propagation of singularities.

In an elliptic point of $S$ we have two directions $(x_1,x_2)$ but in a hyperbolic point 
several cases are possible\\
Let us note,
$k_1=Infk_x <0 < k_2 =Supk_x$ which correspond to the principal curvatives. We have the following results:

\parindent=20pt
{\it 1. If we suppose that  $-k_1 < k_2$ then,

\parindent=40pt
If $\lambda \in ]\ 0, Ek_1^2\ [$, then there are four directions of propagation

If $\lambda \in ]\ Ek_1^2, Ek_2^2\ [$, then we have only two directions.

\parindent=20pt
and if $\lambda > Ek_2^2$, then there are no direction.

2. If  $-k_1 > k_2$ , then there are two directions of propagation  if $\lambda \in ]\ 0, Ek_1^2\ [$ and zero if
$\lambda > Ek_1^2$.}

\bigskip
\parindent=0pt
{\bf Example:}

We consider the hyperbolical paraboloid shell defined by the map,
\begin{equation}
(y^1,y^2) \longrightarrow (x^1 = y^1, x^2 = y^2, x^3 = {c \over 2b}\ [(y^2)^2-(y^1)^2])
\end{equation}
we easily calculate,
\begin{equation}
a_{11}= 1+ {c^2 \over b^4}\  (y^1)^2\ ;\ \ a_{12}=-{c^2 \over b^4} y^1y^2\ ;\ \ 
a_{22}= 1+ {c^2 \over b^4}\  (y^2)^2
\end{equation}
\begin{equation}
b_{11}=-b_{22}= - {c \over b^2 \sqrt {a_{11}a_{22}-a_{12}^2}}\  \ ;\ \ b_{12}=0
\end{equation}
Denoting by $\psi$ the polar angle of the weakness direction $\vec \xi$, 
  $\psi \in ]-{\pi \over 2}, {\pi \over 2}[$, we obtain
\begin{equation}
 \lambda = E.k_x^2  = E \Big[ {b_{11} (tan^2 \psi-1) \over 
A(tan \psi,1) }\ \Big ]^2 
\end{equation}
For $b=0,5$, $c=0,1$ and $y^1=y^2=0$, we have
\begin{equation}
 \lambda = E \Big[ {0,4 (tan^2 \psi-1) \over 
tan^2 \psi+1 }\ \Big ]^2 
\end{equation}
and then
$\Sigma_{ess}= E\  [0, \ 0.16]$. In this case $k_1=-k_2$ and we have four directions (which could coincide) which are symmetrical about the 
polar axis as it is shown on the following figures.
\vfil\eject
For $\lambda = 0$, we have two double directions. This case corresponds to the static problem and the directions of propagation (which are the same that the weakness ones),
are also those of the asymptotic curves of the surface $(b_{11}=-b_{22}\  ;\ \ b_{12}=0)$,

\centerline {\includegraphics{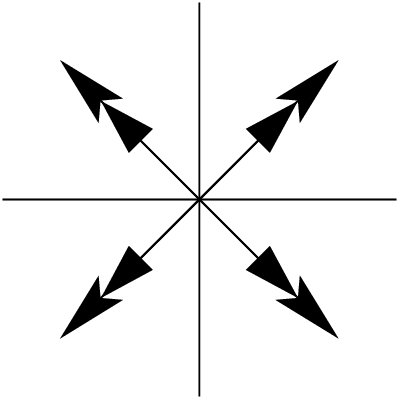}
}

Then when $\lambda$ increases, we obtain four directions, for example for
$\lambda = 0,04.E$,

\centerline {
\includegraphics{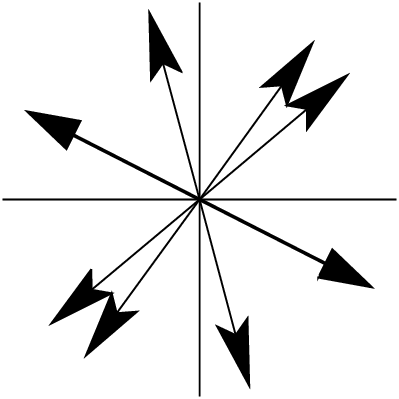}
}
and for
$\lambda = 0,12.E$,

\centerline {
\includegraphics{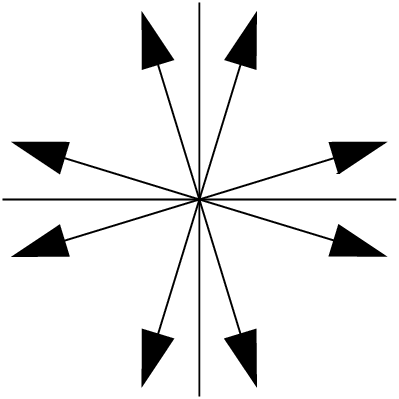}
}
When $\lambda$ reaches the upper bound of the essential spectrum,
$\lambda = 0,16.E$, two directions disappear and the others become coincident:

\centerline {
\includegraphics{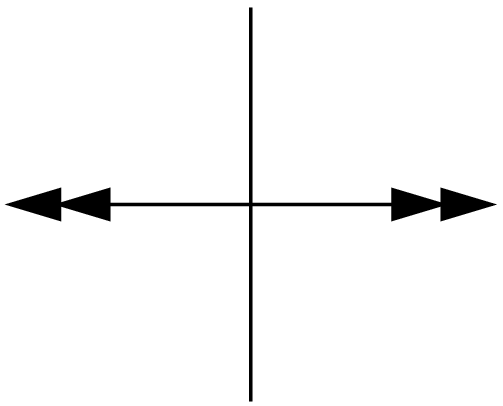}
}

Let us consider another point $y^1=2,5$ and $y^2=5$. We have,
\begin{equation}
 \lambda = E \Big[ {0,2 (tan^2 \psi-1) \over 
3(5tan^2 \psi -4 tan \psi+2) }\ \Big ]^2 
\end{equation}
and $\Sigma_{ess}= E\  [0, \ 2.36\ 10^{-3}]$. Here, $k_1^2= 0.232\ 10^{-3}$ and if
$\lambda \in [0, 0.232\ 10^{-3}E[$, we have four directions:

For $\lambda = 0$ then  $\psi = 45^o$ and we have the two double directions of the asymptotic curves as in the previous case,

\centerline {
\includegraphics{des1.eps}
}

then we obtain four directions. For
$\lambda = 0.1 \ 10^{-3}E$  then  $\psi_1 = -75,2^o$,  $\psi_2 = -26,8^o$,  $\psi_3 = 39,6^o$,
  $\psi_4 = 54^o$,

\centerline {
\includegraphics{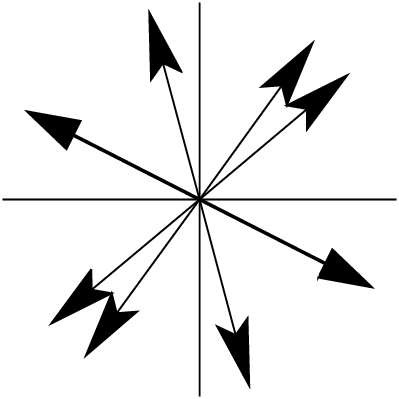}
}

For $\lambda = 0.2\  10^{-3}E$  then  $\psi_1 = 85,3^o$,  $\psi_2 = -19,8^o$,  $\psi_3 = 37,7^o$,
  $\psi_4 = 62,9^o$,

\centerline {
\includegraphics{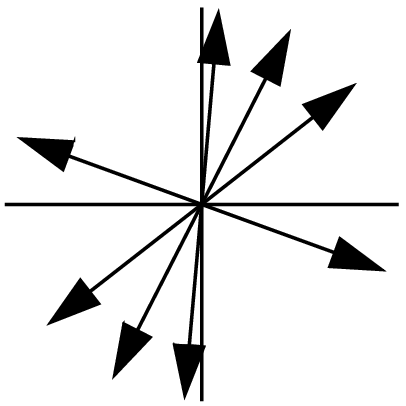}
}

\vfil\eject
For $\lambda = 0.232 \  10^{-3}E=Ek_1^2$,  two directions become coincident in  $\psi_1 = \psi_4 =73^o$ and  $\psi_2 = -18,4^o$,
$\psi_3 = 37,2^o$,

\centerline{
\includegraphics{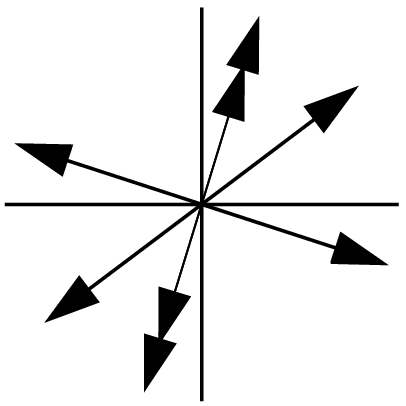}
}

For the values of $\lambda$ larger than $0.232 \  10^{-3}E$, there are only two directions:

If $\lambda = 0.232\  10^{-3}E + 0$ then $\psi_1 = \psi_4 $ disappear,

\centerline {
\includegraphics{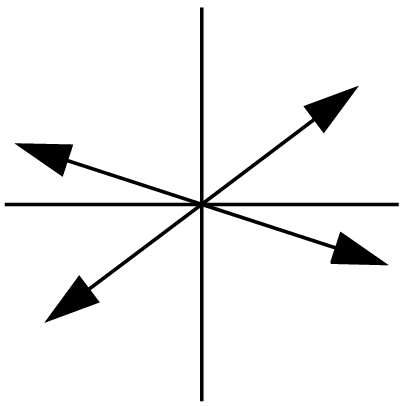}
}

for $\lambda =   10^{-3} E$  then    $\psi_2 = -1,5^o$ and $\psi_3 = 30,5^o$,

\centerline {
\includegraphics{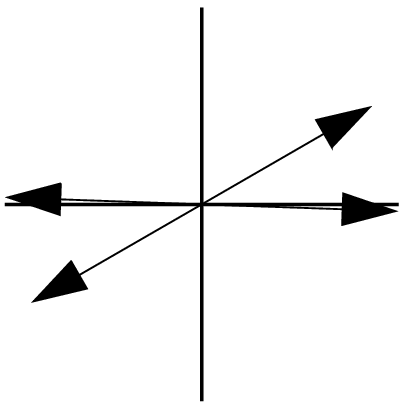}
}

for $\lambda = 2\ 10^{-3}E$  then   $\psi_2 = 10,2^o$,  $\psi_3 = 23,6^o$

\centerline {
\includegraphics{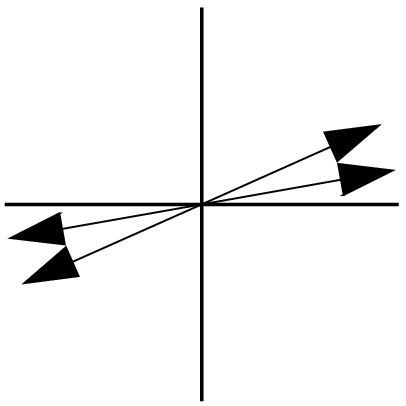}
}

and finally for
$\lambda = 2.36\  10^{-3}E$ the two directions become coincident in  $\psi_2 =\psi_3 = 17,4^o$,

\centerline {
\includegraphics{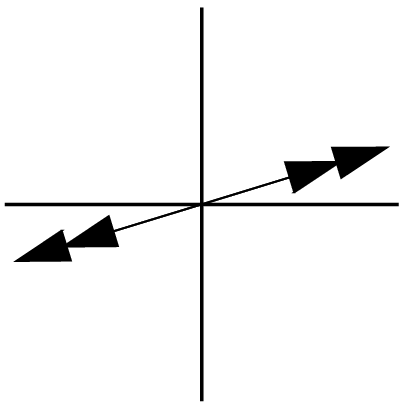}
}
and then they disappear because we go out of the essential spectrum.

\section{Propagation of singularities in a model problem}

The equations of vibrating elastic thin shells in membrane approximation are rather complicated. So to give
a good idea about the properties of propagation of singularities we will first
study a model problem (cf.[13]) and then we will look at the case of a particular shell.\\
We consider the following spectral problem, 
\begin{equation} \label {38}
\left\{
\begin{array}{ccc}
-\Delta u_1 + bu_{2,2} &- \lambda u_1&= f_1  \\
-b u_{1,2} +cu_2 &-  \lambda u_2
 &= f_2 
\end{array} \right.
\end{equation}

where the unknowns $(u_1,u_2)$ are functions of the two variables $y=(y^1,y^2)$ in a domain
$\Omega$ of the plane. We suppose that $b$ and $c$ are given real numbers
 and $f=(f_1,f_2)$
is a right-hand side which will be defined later.
The boundary conditions are for example Dirichlet conditions $u_1=0$ on the boundary
$\partial \Omega$.

\subsection{Essential spectrum}
We define the spaces $H= L^2(\Omega) \times L^2(\Omega)$ and $V=
H^1_0(\Omega) \times L^2(\Omega)$
and  we can write the problem in the form,
 \begin{equation}
 (A-\lambda)u=f 
 \end{equation}
where $A$ is a selfadjoint operator in $H$.\\
The problem involves derivatives of second order in $u_1$ and of first 
 in $u_2$. This problem of mixed order does not satisfy the classical properties of compactness (obviously the inclusion of $V$ in $H$ is not compact). We determine the essential spectrum by writing that this problem is not elliptic in the sense of
Douglis and Nirenberg. We have two equations and two unknowns and the indices are $s_1=1$ and $s_2=0$ (cf.[11 ch. III], [5]). We look for the existence of a nonzero real vector ($\xi_1$, $\xi_2$)
so that the determinant of the principal symbol vanishes,
\begin{equation}\label{36}
det \pmatrix{\xi_1^2+\xi_2^2 & ib\xi_2 \cr
-ib\xi_2 & c-\lambda\cr} =0 
\end{equation}
and it is easily checked that $\lambda$ must belong to  $[c-b^2,c]$. That segment is the essential spectrum which does not depend on the domain $\Omega$.

In general, the set of eigenvalues  of a selfadjoint operator in a Hilbert separable space is denumerable and the corresponding eigenvectors are orthogonal. So, in particular, the eigenvalues of $A$ contained in the essential spectrum have measure zero. 

Let us consider for instance that $\Omega$ 
is the square $[-{\pi \over 2}, {\pi \over 2}]^2$, with Dirichlet conditions on the edge. The functions,
\begin{equation}u_1(y^1,y^2)= A_1cos(2p+1)y^1.cos(2n+1)y^2\end{equation}
\begin{equation}u_2(y^1,y^2)= A_2cos(2p+1)y^1.sin(2n+1)y^2\end{equation}
are eigenfunctions of the problem when the following equation is satisfied:
\begin{equation} \label {45}
{b^2\over c-\lambda}= 1+ {(2p+1)^2\over (2n+1)^2}-{\lambda\over (2n+1)^2}\end{equation}
with
$$b(2n+1)A_1+(c-\lambda)A_2=0$$
Then, when the integers $p$ and $n$ are given, we find two eigenvalues which are at the intersections
of a straight line and a hyperbola (see (\ref {45})). One of them is in $]c-b^2,c[$ and the other is larger than $c$.
For any $p$ and $n$ we can show 
that the subset of these eigenvalues is dense in the essential spectrum and also in $]c, +\infty[$.

\subsection{Propagation of  singularities}
For $\lambda$ belonging to $]c-b^2,c[$, 
there exist two different directions
$(\xi_1,\xi_2)$ which satisfy to (\ref{36}). The orthogonal directions are the two directions of propagation $(cos \theta, sin \theta)$ with,
 \begin{equation}\label{43}
 tan^2 \theta = {\lambda -(c-b^2) \over c-\lambda} 
 \end{equation}
At the extremities of the essential spectrum ($\lambda = c-b^2$, resp. $\lambda = c$), the two directions
of propagation become coincident ($\theta=0$, resp. $ \theta = {\pi \over 2}$).\\
We can note that the substitution of $u_{2,2}$ into the first equation gives the wave equation,
 \begin{equation} \label {42}
 -u_{1,11}+{\lambda -(c-b^2) \over c-\lambda} u_{1,22} -\lambda u_1 = f_1-{b
 \over c-\lambda}f_{2,2}
 \end{equation}
whose characteristic directions correspond to the directions of propagation which
have been defined by $\theta$ (cf.[4]). \\
If $\lambda$ is in the essential spectrum but is not an eigenvalue of $A$, then the range of $A-\lambda$ is dense in
 $V'= H^{-1}(\Omega)\times L^2(\Omega)$ and $A-\lambda$ is injective. For some right-hand sides $f$, 
there exist solutions which are unique but the resolvent is not continuous in these spaces.\\
Let us consider problem (\ref {38}) with the right-hand side $f_1=0$; $f_2=\delta (y^1) \delta(y^2)$.
This right-hand side does not belong to $V'$ but we can consider this problem as the research of
a fundamental solution (cf.[6]). \\
As $\lambda $ is not an eigenvalue, if there exists a solution, then it is unique.\\
When $\lambda$ is exterior to the essential spectrum,  equation (\ref {42}) becomes
 \begin{equation}
 -u_{1,11}+{\lambda -(c-b^2) \over c-\lambda} u_{1,22} -\lambda u_1 = {-b \over c-\lambda}\delta(y^1)\delta'(y^2)
 \end{equation}
 and is elliptic and the 
right-hand side belongs to
$H^s(\Omega)$ with $(s<-2)$ (cf.[10]). If the solution $u_1$ exists, then it belongs to $H^{s+2}(\Omega)$ 
(and $u_2$ to $H^{s+1}(\Omega)$).
In that case there is no propagation.

Let us suppose now that $\lambda$ belongs to $]c-b^2,c[$ and is not an eigenvalue, problem (\ref {38}) can be also written in the form,

\begin{equation}
\left\{ 
\begin{array}{ccc}
-\Delta u_1 + bu_{2,2} &- \lambda u_1 &= 0  \\
-b u_{1,2} +cu_2 &-  \lambda u_2  &=\delta (y^1) \delta(y^2-y^1tan \theta)
\end{array} \right.
\end{equation}

where $tan \theta$ is the slope of one of the characteristics $(D_{\theta})$. We will look for
 solutions in the form of an asymptotic expansion of the singularities across the characteristics. To take orders of derivatives 
in consideration, we propose the following expansions,
\begin{equation}
\left\{ 
\begin{array}{cc}
u_1(y^1,y^2) &= U_1^1(y^1) \delta(y^2-y^1tan \theta) + U_1^2(y^1)Y(y^2-y^1tan \theta) +...  \\
u_2(y^1,y^2) &= U_2^1(y^1) \delta'(y^2-y^1tan \theta) + U_2^2(y^1) \delta(y^2-y^1tan \theta) + ...
\end{array} \right.
\end{equation}

where $Y$ is the Heavyside function. We substitute these expressions in the problem and we can identify
the leading terms, in $\delta''(y^2-y^1tan \theta)$ in the first equation and $\delta'(y^2-y^1tan \theta)$ in the second.\\
We obtain a linear system,
\begin{equation}
\left\{ 
\begin{array}{cc}
(1+tan^2 \theta)U_1^1-bU_2^1&= 0  \\
-bU^1_1+(c-\lambda)U_2^1&= 0
\end{array} \right.
\end{equation}

which admits nonzero solutions because the determinant vanishes when $\lambda$ belongs to the essential spectrum and $\theta$ is defined by (\ref{43}).\\
 The identification of the next terms gives
another linear system,
\begin{equation}
\left\{ 
\begin{array}{cc}
(1+tan^2 \theta)U_1^2(y^1)-bU_2^2(y^1) &= 2 tan \theta  {dU^1_1 \over dy^1}(y^1) \\
-bU^2_1(y^1)+(c-\lambda)U_2^2(y^1) &=\delta(y^1)
\end{array} \right.
\end{equation}

with the same determinant. To obtain solutions, the right-hand sides have to satisfy the following compatibility
condition,
\begin{equation}
 2 (c-\lambda) tan \theta\  {dU^1_1 \over dy^1}(y^1) +b\delta(y^1) =0 
\end{equation}
and then we can find function $U_1^1$ which defines the propagation of the singularity 
along the characteristic $(D_{\theta})$
\begin{equation}\label {52}
 U_1^1(y^1)= {-b \over 2(c-\lambda) tan \theta} (Y(y^1) + C)
\end{equation}
where $C$ is arbitrary and can be obtained by the Dirichlet conditions on the edge of the domain.\\
In that case, we can find a solution $u_1$ of (\ref {38}) belonging to
$H^{s+{3\over2}}(\Omega)$, $(s<-2)$. Then the solution $u_2$ belongs to $H^{s+{1\over2}}(\Omega)$.
The singularities of these solutions are then much more
important than in the case when $\lambda$ is out of the essential spectrum and moreover we
 saw that this singularity can propagate.

\subsection{Reflection of the singularities}

The propagation of singularities is characterized by $U_1^1$ and is along the characteristic straight line which cuts 
the edges of the domain in two points. So we will have two conditions to determine only one constant $C$. 
It shows that the singularity will not disappear by reaching a point $P$ on $\partial \Omega$. 
We will have a reflection on another characteristic the slope of which is $-tan \theta$.

\centerline {
\includegraphics{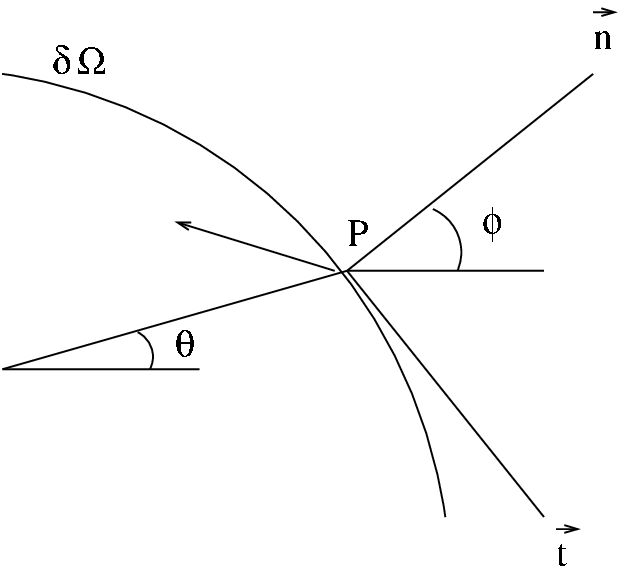}
}

Let $(P,\vec t, \vec n)$ be a local referential. The vector $\vec t$ is tangent to $\partial \Omega$
and $\vec n$ is normal and exterior. We denote by $\phi$ the polar angle of $\vec n$ in the plane $(y^1,y^2)$
and (n,t) are the coordinates of a point in this local referential. By shifting the origin of $y$ in $P$,
we have 
$y^1=tsin \phi +n cos \phi$ and  $y^2=-t cos \phi + nsin \phi$.
We write that the displacement is the superposition of two singularities, incident one $u_1$ and
reflected one $\widehat u_1$, which gives the leading terms on the form
\begin{equation}U_1^1(y^1) \delta(y^2-y^1tan
 \theta) + \widehat U_1^1(y^1)) \delta(y^2+y^1tan
 \theta)
\end{equation}
By using the new coordinates, we obtain on the boundary $\partial \Omega$,
\begin{equation} ({U_1^1(tsin \phi)\over -cos \phi - tan \theta.sin \phi} 
  +{\widehat U_1^1(tsin \phi)\over -cos \phi + tan \theta.sin \phi})  \delta(t)
\end{equation}
which must vanish. Therefore,
\begin{equation}
 {U_1^1(tsin \phi) \over cos(\phi-\theta)} + {\widehat U_1^1(tsin \phi) \over
 cos(\phi+\theta)}=0 
\end{equation}
which shows that if the incident characteristic is almost tangential to the contour
of the domain, then $cos(\phi-\theta)$ is small and the intensity of the reflected singularity is very strong.
In certain cases, the mechanism of propagation and reflection may 
improve the intensity of the singularities.

By introducing the expressions of $U_1^1$ and $\widehat U_1^1$, we easily obtain the following relation,
 \begin{equation}
 C cos(\phi+\theta) - \widehat C cos(\phi-\theta) =2 sin\phi. sin \theta 
 \end{equation}
Obviously the reflected singularity on the characteristic $(D_{-\theta})$ will be reflected again
when this straight line will reach the edge in another point and so on.

As a first example, let us suppose that $\Omega$ is a disk and that $ \theta = {\pi \over 4}$.
By following the characteristics from a point inside $\Omega$, we see that a singularity will be
propagated and reflected along the sides of a rectangle which is inscribed in the circle 
 $\partial \Omega$. 
\medskip

\centerline {
\includegraphics{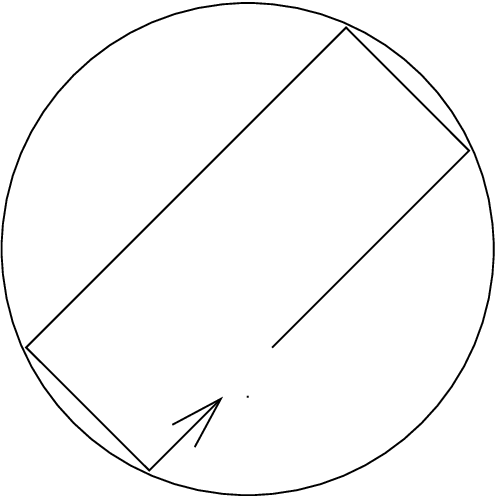}
}

We denote by $C_1$, $C_2$, $C_3$ and $C_4$, the constants corresponding
to the four sides and we have the following system,

\begin{equation}
\left\{ 
\begin{array}{cc}
C_1 cos(\phi+{\pi \over 4})-C_2 cos(\phi-{\pi \over 4}) &= \sqrt2 sin \phi   \\
C_2 cos(\phi-{\pi \over 4})+C_3 cos(\phi+{\pi \over 4}) &= -\sqrt2 cos \phi \\
C_3 cos(\phi+{\pi \over 4})-C_4 cos(\phi-{\pi \over 4}) &= \sqrt2 sin \phi \\ 
C_4 cos(\phi-{\pi \over 4})-C_1 cos(\phi+{\pi \over 4}) &= -\sqrt2 cos \phi  
\end{array} \right.
\end{equation}

The determinant is zero and the compatibility condition is satisfied.

This shows that the asymptotic structure (or at least its leading terms) exists and is not
unique. This suggests that the corresponding value $\lambda = c-{b^2 \over 2}$ is an eigenvalue.
It is certainly the case for $b^2=2c$ because $\lambda$ 
is equal to zero and the homogeneous equation which is associated to (\ref {42}) admits some nonzero
solutions vanishing on $\partial \Omega$.

Let us consider now a second example in which  $\Omega$ 
is the square $[-{\pi \over 2}, {\pi \over 2}]^2$. We may have a lot of successive reflections on the edges and we obtain an angled line.

\centerline {
\includegraphics{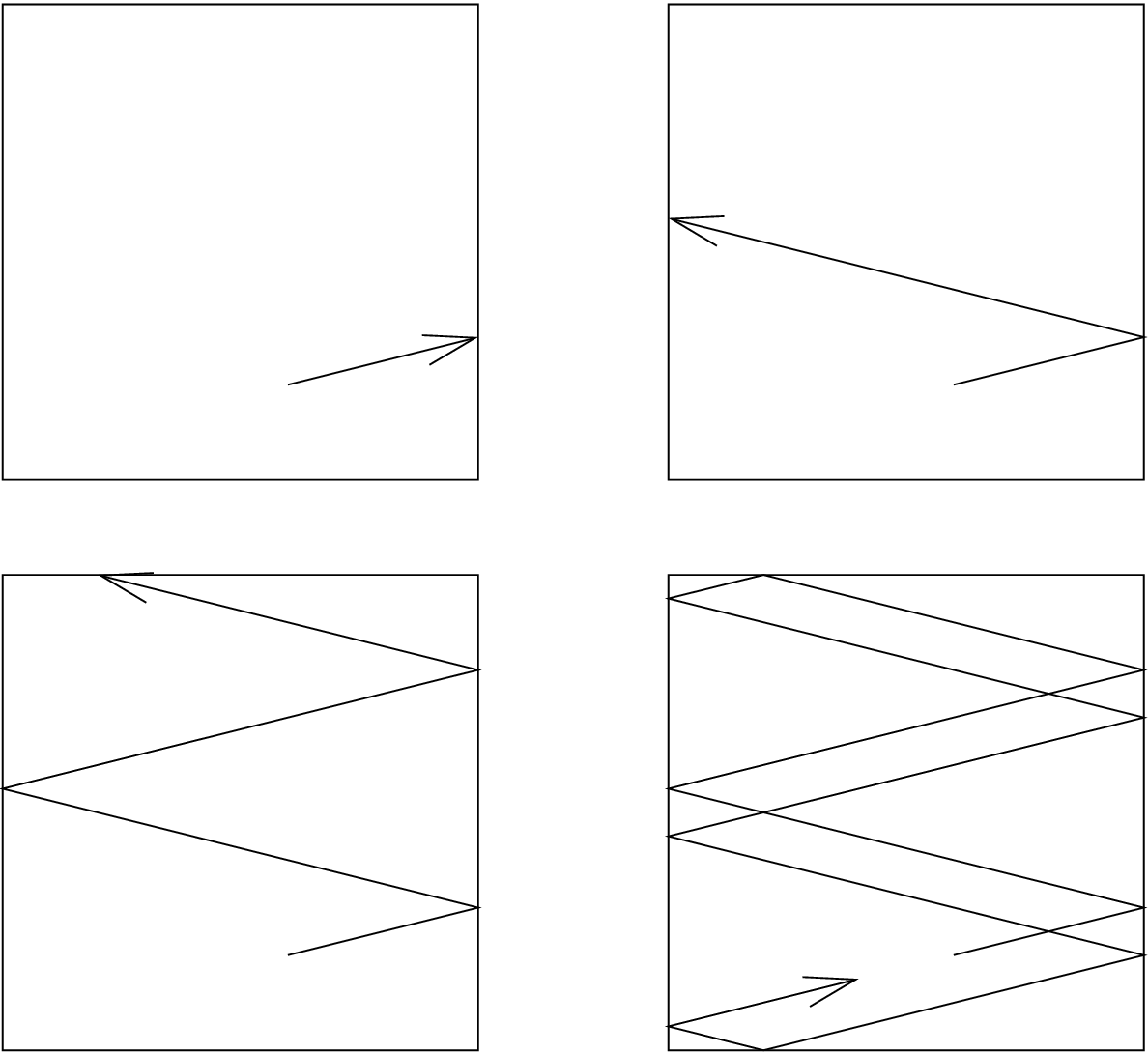}
}

We denote by $C_1$, $C_2$, ... $C_k$, ...  the constants and we have the following system for all value of $ \theta$
different from multiples of ${\pi \over 2}$ ,

\begin{equation}
\left\{ 
\begin{array}{cc}
C_1 -C_2  &= 0   \\
C_2 +C_3  &= -2  \\
C_3 -C_4  &= 0   \\
C_4 +C_5  &= -2  \\
etc...   \\
\end{array} \right.
\end{equation}

It is not hard  to see that the piecewise straight characteristic is closed if and only if $tan\theta$ is a rational number. In this case
we obtain a linear system the determinant of which is zero as in the previous example and the compatibility condition is satisfied.

For other value of $\theta$ (or other $\Omega$) the corresponding trajectory of the reflected
singularities may be somewhat entangled, so implying complicated phenomena of resonance.

Let us remark finally that  at the extremities of the essential spectrum,
($\lambda = c-b^2$ and $\lambda = c$), the two directions of propagation are parallel to the 
axes of coordinates. The incident and reflected directions would be the same so the reflection
does not make sense.

\section {The case of the cylindrical thin shell}

\parindent=0pt
We consider the cylindrical shell defined by the map,
\begin{equation}
(y^1,y^2) \in [0,1] \times [0,2\pi[ \longrightarrow (x^1 = Ry^1, x^2 = R cosy^2, x^3 = R siny^2)
\end{equation}
 We calculate,
\begin{equation}
a_{11}= a_{22}= R^2
\end{equation}
\begin{equation}
b_{22}=R
\end{equation}
The other coefficients of the fundamental forms and the Christoffel coefficients are equal to zero.

 The essential spectrum is the segment $[0, {E \over R^2}]$. Denoting by $\lambda$ a value in the essential spectrum 
 and by $\theta$ the polar angle of the directions of propagation  $\vec x$, 
 we obtain
\begin{equation} \label{62}
 \lambda = {E \over R^2} \Big[ {tan^2 \theta \over 
1 + tan^2 \theta }\ \Big ]^2 
\end{equation}
 
 The equations of the vibrating shell in the membrane approximation are,

\begin{equation}
\left\{ 
\begin{array}{rc}
 T^{11},_1+ T^{12},_2 &=- \lambda u^1 \\
 T^{12},_1+ T^{22},_2 &= -\lambda u^2 \\
RT^{22} &= - \lambda u^3
\end{array} \right.
\end{equation}

where $u^1$, $u^2$ and $u^3$ are the contravariant components of the displacement 
and $T^{\alpha \beta}$, those of the stress tensor. By using the stiffness matrix and the strains $\gamma_{\lambda \mu}$
we have,
\begin{equation} \label {64}
\left\{ 
\begin{array}{rl }
 T^{11} & = K (u_{1,1}+ \nu u_{2,2}-\nu R u_3) \\
T^{22} &= K (\nu u_{1,1} +  u_{2,2}- R u_3)\\
T^{12} &= K {1-\nu \over 2}(u_{1,2} +  u_{2,1})
\end{array}\right.
\end{equation}

where $K= {Eh \over (1-\nu^2)R^4}$. 

We obtain the following homogeneous spectral problem,
\begin{equation}
\left\{ 
\begin{array}{rc}
 u_{1,11} + {1-\nu \over 2}u_{1,22}+ {1+\nu \over 2}u_{2,12}-\nu R u_{3,1}&= - {\lambda \over KR^2} u_1 \\
{1+\nu \over 2}u_{1,12}+ {1-\nu \over 2}u_{2,11} +u_{2,22}- R u_{3,2}&= - {\lambda \over KR^2} u_2 \\
\nu u_{1,1} +  u_{2,2}- R u_3&= - {\lambda \over KR} u_3 
\end{array}\right.
\end{equation}

To study the propagation of the singularities as in section 3, we introduce the following right-hand side which represents a point normal force
\begin{equation}
\left\{ 
\begin{array}{rl}
 u_{1,11} + {1-\nu \over 2}u_{1,22}+ {1+\nu \over 2}u_{2,12}-\nu R u_{3,1}+ {\lambda \over KR^2} u_1&=0 \\
{1+\nu \over 2}u_{1,12}+ {1-\nu \over 2}u_{2,11} +u_{2,22}- R u_{3,2}+ {\lambda \over KR^2} u_2&=0 \\
\nu u_{1,1} +  u_{2,2}- R u_3+ {\lambda \over KR} u_3 &={1\over KR}\delta (y^1) \delta(y^2) \\
\end{array}\right.
\end{equation}

We suppose that  $\lambda $ belongs to the essential spectrum but is not an eigenvalue. 
We search for asymptotic expansions of the displacements on the form,
\begin{equation}
\left\{ 
\begin{array}{rl}
u_{\alpha}(y^1,y^2)& = U_{\alpha}^1(y^1) \delta(y^2-my^1) + U_{\alpha}^2(y^1)Y(y^2-my^1) +... \  \ {\alpha}=1,2 \\
u_3(y^1,y^2) &= U_3^1(y^1) \delta'(y^2-my^1) + U_3^2(y^1) \delta(y^2-my^1) + ...
\end{array} \right.
\end{equation}

where $m = tan \theta$ is the slope of the direction of propagation, and we substitute these expressions in the problem. By identifications of
the leading terms, we obtain the following linear system,
\begin{equation}
\left\{ 
\begin{array}{rl}
 (m^2  + {1-\nu \over 2})U_1^1(y^1)-m {1+\nu \over 2}U_2^1(y^1)+\nu mR U_3^1(y^1)&=0 \\
-m{1+\nu \over 2}U_1^1(y^1)+ ({1-\nu \over 2}m^2 +1)U_2^1(y^1)- R U_3^1(y^1)&=0\\
-\nu m  U_1^1(y^1) +  U_2^1(y^1)+ ({\lambda \over KR}- R) U_3^1(y^1)&=0 
\end{array}\right.
\end{equation}

We shall have nonzero solutions if the determinant of the matrix of this homogeneous system vanishes. 
We then obtain a relation between $m$ and 
$\lambda$,
\begin{equation}
\lambda= KR^2(1-\nu^2){m^4 \over (1+m^2)^2} = {Eh m^4 \over R^2(1+m^2)^2}
\end{equation}

which is the same as (\ref{62}) which was given by the Douglis and Nirenberg method. We note that it does not depend
on the Poisson coefficient. It is easy to calculate $U_1^1(y^1)$ and $U_2^1(y^1)$ according to $U_3^1(y^1)$:
\begin{equation}\label {70}
U_1^1(y^1)= -Rm{\nu m^2-1 \over (1+m^2)^2} U_3^1(y^1)
\end{equation}

and

\begin{equation}\label {72}
U_2^1(y^1)= R{(\nu+2) m^2+1 \over (1+m^2)^2}  U_3^1(y^1)
\end{equation}

Let us write the identifications of the next order terms,
\begin{equation} \label {71}
\left\{ 
\begin{array}{rl}
 (m^2  + {1-\nu \over 2})U_1^2(y^1)-m {1+\nu \over 2}U_2^2(y^1)+\nu mR U_3^2(y^1)
&=2mU_1^1\  '(y^1)- {1+\nu \over 2}U_2^1\  '(y^1)+\nu R U_3^1\  '(y^1) \\
-m{1+\nu \over 2}U_1^2(y^1)+ ({1-\nu \over 2}m^2 +1)U_2^2(y^1)- R U_3^2(y^1)
&=+(1-\nu) m U_2^1\  '(y^1)-{1+\nu \over 2}U_1^1\  '(y^1)\\
-\nu m  U_1^2(y^1) +  U_2^2(y^1)+ ({\lambda \over KR}- R) U_3^2(y^1)&= -\nu U_1^1 \  '(y^1) +{1\over KR}\delta(y^1)
\end{array}\right.
\end{equation}

The determinant of the system is zero and then we have to satisfy a compatibility condition.
Denoting by  $M_{ij}$ the cofactors of the matrix of the linear system and $b_i$ the three right-hand sides of  (\ref{71}),
this condition is,
\begin{equation}
b_1M_{13} -b_2M_{23}+b_3M_{33}=0
\end{equation}

with the same determinant.  A staightforward computation gives the cofactors,
\begin{equation}
\begin{array}{rl}
 M_{13} &= {1-\nu \over 2}( \nu m^3 - m) \\
M_{23} &= {1-\nu \over 2}( (2+\nu) m^2 +1) \\
M_{33} &= {1-\nu \over 2}( m^2+1) ^2
\end{array}
\end{equation}

and then  the right-hand sides have to satisfy the following compatibility
condition,
\begin{eqnarray}
 \lefteqn{ \nu mR (\nu m^2-1) U_3^1\  '(y^1)+{m \over 2}\Big ( (\nu^2+\nu-4) m^2+ 3\nu -1 \Big) U_2^1\  '(y^1)
+{} }\nonumber\\
&&{}  \Big (\nu m^4- {(1+\nu)(2-\nu) \over 2} m^2+ {1-\nu \over 2} \Big) U_1^1\  '(y^1)
 = -(m^2+1)^2 {1\over KR} \delta (y^1)
\end{eqnarray}

We replace $U_2^1(y^1)$ and $U_1^1(y^1)$ by their values (cf. (\ref{70}) and (\ref{72})), and we finally 
obtain
\begin{equation}
4R(1-\nu^2){m^3 \over m^2 +1} U_3^1\  '(y^1)=(m^2+1)^2 {1\over KR} \delta (y^1)
\end{equation}
and then we can find the equation of propagation. By replacing the value of $K$ (cf.(\ref{64})), we have,
\begin{equation}
 U_3^1\  '(y^1)= {R^3 (1+m^2)^3 \over 4Ehm^3}  \delta(y^1) 
\end{equation}
and we see that this equation does not depend on the Poisson coefficient.

The function $U_3^1$ which defines the propagation of the singularity is
\begin{equation}
 U_3^1(y^1)= {R^3 (1+m^2)^3 \over 4Ehm^3} (Y(y^1) + C)
\end{equation}
where $C$ is arbitrary. The form is the same as (\ref{52}) that we obtained in the model problem.
We can also have reflections of the singularities at the intersection with the boundary.

The previous calculus are not valid if $m=tan \theta$ and then $\lambda$ are equal to zero. In that case which is the static case, the propagation of singularities is rather different (see [8], [9]) and there is no reflection (cf. [7]).

That study of the propagation of singularities has been done in a particular case of isotropic shell. The equations of vibration
of a cylindrical shell have constant coefficients and then the propagations are along straight lines. 
The general case is more complicated: \\
The segments which constitute the essential spectrum could be different in every point of the surface $S$. The characteristic curves are some pieces of curves and the propagations along them would depend on the reached point. For example, it is conceivable that the value of the spectral parameter which is given, will be go out of the essential spectrum in some point and that the propagation will stop.\\
Moreover, if some propagation reaches a hyperbolic point at the edge of a shell with four directions of propagation, then we will have several possibilities for the reflection and we do not know what will happen.

\vfil\eject

\noindent{\bf\Large References}\medskip

\noindent [1] M. Bernadou , "Méthodes d'éléments finis pour les problèmes de coques minces"\\
 Masson. Paris. 1994.

\noindent [2] A. Campbell , "A model problem for vibration of thin elastic shells. propagation and reflection of singularities"\\
 C. R. Acad. Sci. Paris. t. 329. S\'erie II b. p.1-5. 2001.

\noindent [3] A. Campbell , "Spectre essentiel et directions de propagation dans les vibrations de coques minces élastiques en approximation membranaire"\\
 Actes du cinquième colloque national en calcul des structures. Giens p.101-106. Mai 2001.

\noindent[4] Yu. Egorov and M. Shubin, "Linear partial differential equations.
Foundations of the classical theory"\\
Encyclopaedia of Mathematical Sciences, Vol. 30, Springer, 1991.

\noindent [5] G. Grubb and G. Geymonat, "The essential spectrum of elliptic systems of mixed orders"\\
Math. Annal. 227 p.247-276. 1977.

\noindent [6] L. Hörmander, "The Analysis of Linear Partial Differential Operators"\\
Springer, Grundlehren, vol. 256, 257, 1983.

\noindent [7] P. Karamian, "R\'eflexion des singularit\'es dans les coques hyperboliques inhib\'ees"\\
 C. R. Acad. Sci. Paris. t. 326. S\'erie II b. p.609-614. 1998.

\noindent [8] P. Karamian, J. Sanchez-Hubert, E. Sanchez-Palencia, "Non-smoothness in the asymptotics of thin shells and propagation of singularities. Hyperbolic case"\\
Int. J. Appl. Math. Comput. Sci. Vol. 12, No 1, p.81-90. 2002.

\noindent [9] P. Karamian, J. Sanchez-Hubert, E. Sanchez-Palencia, "Propagation of singularities and structure of layers in shells. Hyperbolic case"\\
 Computers and Structures. p.747-768. 2002.

\noindent [10] J.L. Lions et E. Magenes, "Probl\`emes aux limites non homog\`enes et applications"\\
 Vol.1, Dunod, Paris, 1968.

\noindent [11] J. Sanchez-Hubert and E. Sanchez-Palencia, "Coques \'elastiques minces, propri\'et\'es
asymptotiques" \\
 Masson, Paris, 1997.

\noindent [12] J. Sanchez-Hubert and E. Sanchez-Palencia, "Vibration and coupling of
continuous systems"\\
 Springer, Berlin, 1989.
 
\noindent [13] E. Sanchez-Palencia and D. Vassiliev, "Remarks on vibration of thin elastic shells
and their numerical computation"\\
C. R. Acad. Sci. Paris. t. 314. SÈrie II. p.445-452. 1992.

 \end{document}